# Enhancing Urban Traffic Safety: An Evaluation of Taipei's Neighborhood Traffic Environment Improvement Program

Frank Yao Huang  
National Taiwan University

Po-Chun Huang  
National Chengchi University

## Abstract

In densely populated urban areas, where interactions between pedestrians, vehicles, and motorcycles are frequent and complex, traffic safety is a critical concern. This paper evaluates the Neighborhood Traffic Environment Improvement Program in Taipei, which involved painting pedestrian walkways green, adjusting no-parking red/yellow lines, and painting speed limit and stop/slow signs on lanes and alleys. Exploiting the staggered rollout of the program's implementation and administrative traffic accident data, we found that the program reduced daytime traffic accidents by 5% and injuries by 8%, while having no significant impact on nighttime incidents. The effectiveness of the program during the day is mainly attributed to the green-painted walkways, with adequate daytime lighting playing a part in the program's effects. Our findings indicate that cost-effective strategies like green-painted pedestrian lanes can be effective in areas with dense populations and high motorcycle traffic, as they improve safety by encouraging pedestrians to use marked areas and deterring vehicles from entering these zones.

Keywords: Neighborhood Traffic Environment Improvement Program, Traffic Safety, Poisson Regression, Difference-in-Differences

# 1. Introduction

The World Health Organization (WHO) reports that worldwide, road accidents claim approximately 1.3 million lives annually and result in tens of millions of non-fatal injuries.[1] In densely populated urban areas, where interactions between pedestrians, vehicles, and motorcycles are frequent and complex, traffic safety is a critical concern. Effective traffic control policies play an important role in reducing accidents, thereby saving lives and cutting healthcare costs. Cities around the world have responded by implementing various traffic control measures, such as the construction of sidewalks and crosswalks, installation of speed limit and stop/slow signs, and traffic calming devices like road humps. More than just legal regulations, these measures also serve as essential strategies for accident prevention. However, little is known as to whether these policies result in positive outcomes (Duncan, 2023).

This paper evaluates the Neighborhood Traffic Environment Improvement Program (NTEIP) in Taipei, Taiwan.[2] Taipei's urban landscape is marked by its narrow alleys and high density of people and vehicles, particularly motorcycles. This setting poses distinct challenges for traffic safety, requiring innovative and effective solutions.

Launched in August 2015 by the Taipei City Government in response to these challenges, the NTEIP implemented several strategic measures, such as painting pedestrian lanes on narrow roads green, serving as a visual nudge for pedestrians to walk within these marked areas and for drivers to keep vehicles out of these zones. The NTEIP also involved adjusting no-parking zones to alleviate congestion and painting speed limit and stop/slow signs in alleys to control vehicle speeds.³ These initiatives collectively aimed to minimize accidents and bolster pedestrian safety in Taipei's densely populated areas.

---

[1] WHO global status report on road safety 2021: https://www.afro.who.int/publications/global-status-report-road-safety-time-action.
[2] In Taiwan, the Ministry of Health and Welfare's 2020 statistics showed that accidents ranked sixth among the top ten causes of death. Of these, transportation accidents accounted for 46.1% of accidental injuries, a significant proportion.

However, the impact of the NTEIP on traffic safety is not straightforward. While initiatives such as painted pedestrian lanes and adjusted no-parking zones aim to reduce illegal parking and increase road space, they might inadvertently lead to higher driving speeds, potentially escalating both the risk and severity of accidents. Moreover, while intended to enhance safety, visual cues such as speed limit or stop/slow signs could distract drivers and inadvertently contribute to accidents and casualties, as suggested by Beijer et al. (2004).

In this analysis, we used administrative data for traffic accidents and exploited the staggered implementation of the NTEIP to identify the causal effect of these measures on traffic safety. Assuming roads that were treated under the NTEIP and roads that had not yet been treated would have similar trends in the absence of the program, the divergence in accident and injury rates between them after the NETIP's implementation can be attributed to the program's effects. Our event study analysis suggests that the treatment and control groups were indeed on similar trajectories before the NTEIP was rolled out, validating the assumption of our difference-in-differences (DID) design.

We focused on two primary metrics: the overall number of road accidents and the injuries resulting from car accidents. Given that these figures are count data, represented by nonnegative integers, we utilize Poisson regression integrated with road and time fixed effects to discern the effects of the program. Poisson regression presents clear advantages. On one hand, Roth and Chen (2023) emphasized that, in contrast to models using log(Y+1) where coefficient interpretation can be challenging, the coefficients in a Poisson regression are interpretable as semi-elasticities. This means that they represent the percentage changes in car accidents and injuries attributable to the NTEIP. On the other hand, Wooldridge (2023) highlighted that the linear parallel trend assumption in DID can be questionable when outcomes are limited in range, and the combination of Poisson regression with DID is a more suitable approach when dealing with count data.

---

³ The NTEIP not only implemented the green-painted pedestrian walkways, installed speed limit and stop/slow signs, and adjusted the red/yellow no-parking lines, but also adjusted the car and motorcycle parking grid, motorcycles' exit sidewalks, bicycle parking spaces, etc.. The latter group, though, are less related to road traffic safety, and are therefore not considered in this study.

Our empirical analyses of the effects of the NTEIP measures implemented in the lanes and alleys of Taipei City from 2015 to 2019 yielded several findings. The estimated effect of the NTEIP on overall traffic accidents indicates an insignificant decrease of around 3.5%. However, the effect of the NTEIP on traffic injuries is pronounced, significantly reducing traffic injuries by about 7%, primarily due to the painted-green pedestrian walkways. Furthermore, our estimates suggest that the NTEIP significantly reduces daytime traffic accidents by about 5% and injuries by approximately 8% but has no effect on nighttime accidents or injuries. This pattern indicates that the markings and signs are most effective during daylight, with their visibility enhanced by natural light. Therefore, enhancing lighting in lanes and alleys could be a strategic move to improve nighttime road safety.[4]

This paper contributes to the existing literature in two ways. First, it adds to research evaluating comprehensive pedestrian plans to improve traffic safety in neighborhoods, a comparatively underexplored area. Inada et al. (2020) estimated the effect of Japan's "Zone 30" policy, which introduced 30 km/h zones in 2011. Their estimates using interrupted time series data suggested that the policy had prevented 1,704 cyclist and pedestrian injuries by 2016. Similarly, Seya et al.'s (2021) estimates using propensity score matching suggested that a 30 km/h zone reduced the number of serious injuries when combined with physical devices. In addition, Duncan (2023) used the American Community Survey to show that while there is no significant change in walk commuting after the initial pedestrian plan, modest but significant increases may occur with frequent updates. Our study contributes to this literature by using a natural experiment from the implementation of the NTEIP, demonstrating that measures like green-painted walkways can effectively improve traffic safety in densely populated areas.

---

[4] Under different conditions, drivers' visual cognition or recognition of road signs varies (Bullough, 2017). Zwahlen and Schnell (1999) and Zwahlen et al. (1991) found that during the day, drivers' recognition distances for traffic signals are higher than at night (about 1.2 to 1.8 times greater). Rumar and Ost (1974) observed that dirty traffic signs at night, or in poor lighting conditions, have decreased recognizability, but this effect is limited during the day or under proper lighting. Madleňák et al. (2018) measured that drivers had 38.5% fewer 'fixations' on traffic signs under low-light conditions compared to well-lit conditions. These studies highlight that road users' feedback from traffic signs varies under different lighting conditions.

Second, this paper advances the broader literature on the effectiveness of policy solutions aimed at reducing traffic accidents and injuries. Previous research has indicated that higher speed limits lead to increased driving speeds and greater accident severity (Wilmot and Khanal, 1999; Malyshkina and Mannering, 2008; Ashenfelter and Greenstone, 2004; Van Benthem, 2015), while decreasing speed limits can significantly reduce traffic accidents (Hess and Polak, 2003; Ang et al., 2020). Furthermore, laws mandating the use of seatbelts and helmets have been shown to protect drivers from traffic fatalities (Carpenter and Stehr, 2008; Cohen and Einav, 2003; Dee, 2009; Dickert-Conlin et al., 2011; Blanco et al., 2021). The government can curb speeding not only through legislation, but also by installing traffic signals that remind drivers to decelerate in dangerous areas (Ezeibe et al., 2019; Hijar et al., 2003; Van Houten et al., 1985; Van Houten, 1988). However, there is a concern that these signals might introduce a "distraction effect" (Lauer and McMonagle, 1955; Molino et al., 2009; Beijer et al., 2004), leading to mixed impacts on driver behavior. By providing new evidence on the effectiveness of traffic control policies in neighborhood settings, this paper enriches the ongoing dialogue on urban traffic management and public health.

The structure of the paper is as follows. Section 2 introduces the NTEIP. Section 3 describes the data and sample. We discuss our empirical design in Section 4 and report the estimation results in Section 5. Section 6 concludes.

## 2. Policy Background

Taipei is a city marked by its dense population and narrow alleys that are frequently used for parking. To enhance pedestrian safety and effectively manage vehicle flow in these compact urban spaces, the Taipei City Government initiated the NTEIP in August 2015. Safety enhancement measures introduced under the NTEIP included (1) painting pedestrian paths green, (2) adjusting red/yellow no-parking lines, (3) installing 30 km/h speed limit and stop/slow signs, and (4) other measures: addition or removal of bicycle parking areas, vehicle parking spaces, and motorcycle exits from covered walkways. In our analyses, we focused on the first three improvement measures because the other measures are less related to traffic safety.

The NTEIP measures were strategically implemented in streets, lanes, and alleys across the city, with a particular focus on those less than 12 meters wide. They were also implemented in a staggered fashion over time, which created a valuable natural experiment through which to investigate their effects. Figure 1 in the paper visually illustrates the interventions of interest, showing a stop sign, a slow sign, a 30 km/h speed limit sign, and a green-painted pedestrian path.

The introduction of the NTEIP was a response to several critical urban challenges. First, the lack of early regulations for sidewalks resulted in alleys becoming frequent sites of pedestrian–vehicle conflicts. To address this, the NTEIP had green pedestrian paths painted on roads less than 12 meters wide, which traditionally lacked sidewalks or arcades. This design visually segregated the road, turning one side into green-painted pedestrian paths and leaving the other for vehicle parking, aiming to resolve the space conflict without the need for physical sidewalks.

Second, the program addressed the issue of speeding in lanes and alleys. Despite the existing 30 km/h speed limit, the absence of clear signs often led to vehicles exceeding safe speeds. The installation of "stop," "slow down," and "30 km/h speed limit" signs as part of the NTEIP provided visual reminders for drivers to moderate their speed, prioritizing pedestrian safety.[5]

Third, the NTEIP reevaluated and reorganized parking within Taipei's lanes and alleys. To insure adequate temporary parking for trucks and cars, the arrangement of no-parking red/yellow lines was revised. Red lines in certain areas were converted to yellow to increase legal temporary parking availability. Moreover, in broader alleys, the length of the no-parking red line at the entrance was reduced, creating additional parking spaces for motorcycles without compromising drivers' visibility.

From 2015 to 2020, the NTEIP was fully implemented across all of Taipei's neighborhoods. Out of 456 neighborhoods, 379 (83.11%) were rated exemplary, with over 80% of their alleys (less than 12 meters wide) fully improved. According to 2018 statistics from the Taipei City Government, in the 60 neighborhoods where the project was fully implemented for at least a year, with a completion rate of over 80%, there was a 68% decrease in traffic accidents and a 64% reduction in casualties compared to the previous year.[6]

# 3. Data

---

[5] The speed limit of the main arterial roads in Taipei is 50 km/h, and some roads are higher or lower than this speed limit; details can be found in the road speed limit table produced by the Department of Transportation: https://www.bote.gov.taipei/cp.aspx?n=471C0AA3AAD97ADB.

[6] The estimate came from the Department of Transportation of the Taipei City Government: https://www.dot.gov.taipei/News_Content.aspx?n=C41A7FC0570A20B3&sms=72544237BBE4C5F6&s=62CA36F207FF09F3.

## 3.1 Data for the NTEIP

Our sample consists of the roads that were designated to receive some form of treatment under the NTEIP. We obtained information about which roads these were, the planned timing of the program implementation (from 2015 to 2019), and whether the roads actually underwent the planned improvements from the Taipei City Government Transportation Bureau's website. This website records the year in which NTEIP initiatives were planned for each village in Taipei, as well as the roads in each village intended to receive treatment under those plans.[7]

For the roads selected to receive treatment under the NTEIP, we have information on their planned improvements and locations. These details are meticulously documented, outlining program elements such as the addition of green-painted pedestrian walkways, changes to red/yellow no-parking lines, and adjustments in vehicle and motorcycle parking spaces, as well as installation of speed limit and stop/slow signs. We also obtain the completeness of each road—regardless of whether the improvements contained in the NTEIP were actually implemented on that road.

For the purposes of our study, a road is considered "complete" if it met the improvement requirements of the NTEIP when the program ended. In other words, a road that had already undergone all the improvements before the program began is still considered "complete," even if the plan did not schedule any new improvements for it. Similarly, if some parts of a road were not yet improved at the time of the program's planning, but were completed after the plan was implemented, it is also considered "complete." In contrast, if some parts of a road were not improved before the program and no improvements were planned (e.g., due to private land ownership or roads that were too narrow to make improvements on), then that road is considered "incomplete."

---

[7] For detailed information about the NTEIP, visit: http://111.235.214.55/NeborhodTraficImprove/.

In short, the NTEIP's scope included roads that already met the desired standard at the outset of planning, those that were enhanced during the course of the program's implementation, and those that did not receive any enhancements for various reasons. This study concentrates on three critical improvements related to road traffic safety: green-painted pedestrian walkways, red/yellow no-parking lines, and speed limit and stop/slow signs.

According to Taipei's naming conventions, roadways are designated according to width: broader thoroughfares are called "avenues," "roads," or "streets," while narrower paths are referred to as "lanes" or "alleys." The NTEIP mainly targeted lanes and alleys, although it also included a few broader roads. Improvements on larger roads were expected to cause minimal disruption to traffic, in contrast to changes in narrower alleys, which in some cases significantly impacted traffic flow.

We assigned roads to one of four categories based on the NTEIP criteria: those that already met the standards during the initial planning, those that were improved post-planning but did not include the three improvement features we focused on, those that received at least one of the three critical improvements, and roads that did not meet standards but received no enhancements for various reasons. The first two categories we labeled as "always treated," the third as the "treatment group," and the fourth as "never treated." The appendix provides a detailed breakdown of these categories.

In this study, our focus was on 6,856 lanes and alleys included within the NTEIP, which spanned 453 neighborhoods across 12 districts. These lanes and alleys were divided into three groups: "always treated" (1,891, 27.58%); "treatment" (3,414, 49.50%); and "never treated" (1,551, 22.62%).[8]

---

[8] The planned time for improvements under the program was determined based on individual villages, and some roads that span across different villages. Additionally, there were a few roads where the program was piloted in advance. To clearly define the time of planning for each road, this article excludes the aforementioned two types of roads from the sample.

## 3.2 Data on Traffic Accidents

Data on traffic accidents were obtained from the Taipei City Police Department, and cover the period from 2013 to 2020. The data classifies accidents as A1 (deaths), A2 (injuries), and A3 (vehicle damage). Using information on the time of occurrence, we categorized the accidents into day- and nighttime accidents to understand how lighting conditions influence driver behavior. We matched this traffic accident data with planning and implementation information from the NTEIP to construct annual panel data comprising traffic accidents, related injuries (including deaths), and program information at the road level.[9] In cases where an accident occurs at the junction of two roads, both road names are noted, attributing one accident count to each road. Injury counts were similarly distributed across the two roads.

Table 3 shows summary statistics for traffic accidents and injuries occurring on Types I, II, and III roads. Lanes and alleys historically experience fewer accidents than main roads, often with fewer injuries. Notably, several lanes and alleys reported zero accidents in certain years.[10] On average, as shown in Table 3, each lane or alley recorded slightly more than one accident or injury per year, with a peak of 53 accidents and 41 injuries in any given lane or alley. Nighttime accidents and injuries are less frequent than daytime ones.

# 4. Empirical Design

## 4.1 Poisson Difference-in-Differences

We estimated the following Poisson regression to measure the effect of the NTEIP

---

[9] Deaths are very rare in lanes and alleys. The annual incidence of traffic-related deaths in the lanes and alleys considered in this study was less than 0.2 percentage points. Therefore, we combined traffic deaths with traffic injuries for our analysis.
[10] Nearly half of the 54,848 samples (27,289) have zero traffic accidents in the lanes and alleys, and 34,516 samples have zero injuries.

on the number of traffic accidents and casualties:

$$\log(E(y_{it})) = \alpha_i + \mu_t + \gamma^{DID} D_{it} + \beta X_{it}, \qquad (1)$$

in which $y_{it}$ represents the number of accidents or injuries on road $i$ in year $t$. The dummy variable $D_{it}$ is set to 1 if road $i$ has implemented at least one of the three critical improvements (green-pained walkway, no-parking red or yellow lines, and speed limit and stop/slow signs) by year $t$, and 0 otherwise. Road fixed effects ($\alpha_i$) captures time-invariant characteristics specific to each road, such as width and surrounding environment, which can influence accident or injury rates. Time fixed effects ($\mu_t$) accounts for yearly variations affecting all roads, such as changes in traffic behavior, pedestrian practices, or shifts in road safety awareness due to campaigns or new laws. Control variables $X_{it}$ include population density, vehicle density, and education level in each district of Taipei.[11] To account for serial correlations over years at the road level, standard errors are clustered by road.

$\gamma^{DID}$ is the coefficient of interest—it represents the change in traffic accidents post-NTEIP implementation compared to roads that did not undergo any improvements. In line with recent developments in DID design (e.g., Goodman-Bacon, 2021; de Chaisemartin and D'Haultfœuille, 2020; Roth, 2023), we excluded always treated roads from our estimation sample. We argue, however, that never treated roads do not constitute an appropriate control group. As seen in Table 3, never-treated roads not only had substantially lower accident rates but also significantly fewer traffic injuries compared to treated roads. Therefore, our model (1) forms a staggered DID design that compares roads that underwent the NTEIP's enhancement measures in different years.

---

[11] Owing to the challenge of collecting road-level control variables, we have incorporated district-level variables from the Taipei Statistical Database query system (https://statdb.dbas.gov.taipei/pxweb2007-tp/dialog/statfile9.asp). Population density is calculated based on the district's population per hectare, vehicle density is determined by the number of registered motor vehicles per hectare, and educational level is gauged as the percentage of the population aged 18 years and older with college degrees.

Recent developments in DID design have identified complications with staggered DID designs, particularly when there is variation in treatment timing (Goodman-Bacon, 2021). Specifically, estimates from staggered DID designs can be biased if the control group includes already-treated units. To address this issue, we applied the estimator from Callaway and Sant'Anna (2021) (CSDID), which is robust to variations in treatment timing, as a robustness check in Section 5.5.

## 4.2 Event Study

The DID design relies on the assumption that the number of car accidents and injuries would have followed the same trend for the treatment and control roads in the absence of NTEIP. To investigate this identifying assumption, we consider the following event study specification:

$$\log(E(y_{it})) = \alpha_{i_i} + \mu_t + \gamma_{-3}D_{-3,it} + \gamma_{-2}D_{-2,it} + \gamma_0 D_{0,it} + .. + \gamma_3 D_{3,it} + \beta X_{it} \quad (2)$$

in which $D_{k,it}$ is equal to 1, if road $i$ implemented at least one of the three critical improvements $k$ years before year $t$, otherwise it is 0 ($k=-2,..,2$). The dummy variable $D_{-3,it}$ is an indicator that road $i$ became treated 3 years or more before year $t$, $D_{3,it}$ indicates road $i$ has been treated for 3 years or more in year $t$. We used the year before the program was implemented as the reference period. Thus, the estimated coefficients $\gamma_k$ reflect the effects of the NTEIP relative to one year prior to the program beginning.

Interpreting $\gamma_k$ as causal effects of NTEIP requires the assumption that trends in outcomes of treated roads would have paralleled those of untreated roads in the absence of NTEIP. $\gamma_{-3}$ and $\gamma_{-2}$ are useful for investigating this assumption—they should not be significantly different from zero if treatment and control groups experienced similar outcome trends before the reference year. Given parallel trends prior to the reference year, we can therefore interpret $\gamma_0$ to $\gamma_3$ as the effects of the NTEIP.

## 4.3 Effects of Specific Improving Content

The NTEIP interventions include the implementation of green-painted walkways, the adjustment of red/yellow no-parking lines, and the introduction of speed limit and stop/slow signs. To distinguish the specific effects of each improvement measure and understand their separate contributions to enhancing road safety, we revised model (1) as:

$$\log(E(y_{it})) = \alpha_i + \mu_t + \sum_{k \in \text{Cotent}} \gamma_k D_{it}^k + \beta X_{it}, \qquad (3)$$

where $\sum_{k \in \text{Content}} \gamma_k D_{it}^k$ incorporates the three aforementioned road improvements. The coefficient $\gamma_k$ of each implementation content measures its impact on road traffic safety. For example, when considering the implementation of green-painted walkways, the dummy variable can be defined as:

$$D_{it}^{\text{Walkway}} = 1\{\text{Green walkway}_i\} \cdot 1\{\text{the program has been implemented in year t}\},$$

and $\gamma_{\text{green walkway}}$ represents the effect of green-painted walkways on traffic safety.

The next section presents the estimation results for Models (1)–(3), focusing on the effects of the NTEIP on traffic accidents and injuries. We will differentiate between incidents that occur during daytime and nighttime for each model.

## 5. Empirical Results

## 5.1 Overall Traffic Accidents and Injuries

Table 4 presents the estimated effects of the NTEIP on traffic safety using Model (1). Columns 1 and 2 present the estimated program effects on traffic accidents. Column 1 suggests that the NTEIP decreases traffic accidents by 3.51%, but the estimated coefficient is not statistically significantly different from zero. Using population density, vehicle density, and the proportion of college graduates as controls does not significantly change the results, as seen in Column 2.

Columns 3 and 4 of Table 4 present the estimates for the effects of the NTEIP on the number of people injured in accidents.[12] We can see that the NTEIP is estimated to reduce the number of traffic injuries by approximately 6.55%. After controlling for population density, vehicle density, and education level, the estimate increases to a reduction of 6.98% in traffic injuries.

## 5.2 Daytime and Nighttime Traffic Accidents and Injuries

Tables 5 and 6 investigate whether the presence of sufficient lighting affects the NTEIP's enhancement measures' effectiveness in reducing traffic accidents and injuries. Table 5 changes the dependent variable to the number of daytime traffic accidents and injuries, with all sample and control variables remaining the same as in Table 4. Columns 1 and 2 suggest that the program reduced the number of accidents by about 4.9%. On the other hand, columns 3 and 4 show that the NTEIP reduces daytime traffic injuries by 8–9%. The fact that the decrease in traffic injuries is larger than that in traffic accidents suggests that the NTEIP not only markedly reduces the likelihood of accidents, but also the severity of accidents.

Comparing Tables 5 and 4, the drop in the number of traffic accidents and injuries during daytime accidents is larger than the overall decrease in the number of traffic accidents (3–4%) and injuries (6–7%). This suggests that sufficient lighting makes it easier for road users to recognize road signs, reflecting the substantial impact of the NTEIP on the number of traffic accidents and injuries.

---

[12] Note that if the dependent variable for a particular road is zero for all years, that road does not contribute to the estimation in the Poisson fixed-effects model (Wooldridge, 2010). Therefore, although Columns 3 and 4 use the same sample as those in Columns 1 and 2, the number of observations changes due to the difference in the dependent variable.

Table 6 estimates the effects of the NTEIP on nighttime traffic accidents and injuries; both effects are small and statistically insignificant. This indicates that the program's impact on nighttime road safety is not substantial. This contrasts sharply with the substantial reductions in daytime accidents and injuries reported in Table 5, likely due to inadequate lighting at night. This insufficiency hinders the visibility of road markings and signs, compromising the program's effectiveness in improving road awareness. These results underscore the importance of sufficient lighting in enhancing the visibility of traffic signs and, consequently, improving traffic safety.

## 5.3 Dynamic Effects of NTEIP

This section examines the dynamic effects of the NTEIP on the number of traffic accidents and injuries using the model specified in Equation (3). The empirical results are presented in Table 7. This analysis investigates whether there were pre-program trend differences between the treatment and control groups to examine the credibility of the DID estimates; it also examines the time-varying effects of the program post-implementation.

Table 7 shows the estimates for the overall, daytime, and nighttime number of accidents and injuries with the control variables included. The results suggest that there were no anticipated effects of the program two or three years before implementation. This indicates no significant trend differences in the number of accidents or injuries between the treatment and control groups, supporting the credibility of the estimates in Tables 4–6. In addition, the fact that the program took time to be completed is reflected in the smaller, insignificant decrease in the number of accidents and injuries in the year when improvement measures were completed.

Columns 1 and 2 reveal an overall decrease in the total number of accidents and injuries post-implementation (although the decrease in traffic accidents is insignificant), aligning with the results in Table 5. Columns 3 and 4 indicate a significant decrease in daytime accidents (8.52%) and injuries (15.6%) in the first full year after the enhancement measure was implemented. In the years following the program's implementation, the coefficients are similar in magnitude to those of the first year but less precisely estimated. This might be because the enhancements in some lanes and alleys were completed later in the sample period, and thus we did not observe the effects for a full two or three years after the program was implemented. Lastly, columns 5 and 6 show that the effects on nighttime accidents and injuries post-implementation are small and statistically insignificant.

## 5.4 Effects of Specific Improving Content

Based on model (3), we estimated the effects of three measures implemented under the NTEIP: painting pedestrian walkways green, adjusting the placement of red/yellow no-parking lines, and placing speed limit and stop/slow signage. The empirical results of each improvement in overall, daytime, and nighttime traffic accidents and injuries are presented in Table 8.

Columns 1 and 2 of Table 8 indicate that the individual measures of the NTEIP program have no significant impact on the overall number of traffic accidents and injuries. While green-painted pedestrian walkways slightly reduce both, the decrease is not statistically significant. The adjustments to red/yellow no-parking lines show negative but insignificant effects, and the installation of speed limit and stop/slow signs slightly increases the number of accidents, though not significantly.

Columns 3 and 4 show that during the day, green-painted pedestrian walkways significantly decrease traffic injuries by 11%, clearly demonstrating their role in making drivers more aware of pedestrians. However, adjustments to red/yellow no-parking lines had no significant effect in daytime. Interestingly, speed limit and stop/slow signs, which show negative coefficients, suggest that they do not contribute to the daytime increase in accidents. Across all measures, there is a notable trend: a larger reduction in injuries than in accidents, indicating that the markings and signs installed under the NTEIP are more effective in reducing accident severity than frequency.

At night, according to columns 5 and 6, although imprecisely estimated, the estimated impact of green-painted pedestrian walkways on accidents is positive at 6%. This might be due to poor visibility at night, causing drivers to misinterpret the walkways and leading pedestrians to feel falsely secure, potentially increasing the likelihood of accidents. Adjustments to no-parking lines reduce accidents by 2%, hinting at the benefits of less illegal parking, although this change is not statistically significant. speed limit and stop/slow signs are linked to a 9% increase in accidents and injuries. Although these estimates are imprecise, it is possible that poor night visibility leads drivers to misread the signs, causing distraction and increasing road risk.

## 5.5 Robustness Check

This section presents robustness checks for our main results: that implementation of the NTEIP has significantly reduced daytime traffic accidents and injuries while having no effects at night.

On one hand, we differentiate between A1/A2 and A3 traffic accidents, with A1/A2 involving fatalities or injuries, and A3 involving only vehicle damage. If the NTEIP can reduce accidents by encouraging pedestrians to use green-painted walkways and discouraging vehicles from entering these zones, we would expect to see a reduction in A1/A2 accidents and A3 remaining unaffected. Columns 1–3 of Table 9 demonstrate that the NTEIP measures effectively reduced A1/A2 accidents; overall incidents decreased by 6.46%, and daytime accidents by approximately 7%, without significantly impacting nighttime A1/A2 accidents. Notably, this reduction in daytime A1/A2 accidents also corresponds with an 8–9% decrease in daytime injuries, highlighting the NTEIP measures' effectiveness in reducing accident-related injuries. On the other hand, NTEIP implementation does not appear to affect A3 accidents (columns 4–6), which serves as a falsification test supporting the main findings.

On the other hand, to address concerns about potential biases due to variation in treatment timing, we used the CSDID estimator to avoid using already-treated units as a control group (Goodman-Bacon, 2021). In Table 10, Panel A normalizes traffic accidents and injuries to their 2015 village means for better comparability with the results from Poisson regression. These findings align with our main results, indicating the NTEIP measures' effectiveness in reducing daytime traffic accidents and injuries by a notable margin without impacting nighttime incidents. Specifically, there is a 22% decrease in traffic accidents (1.1 out of 5.1 percentage points) and an 18% decrease in injuries (0.9 out of 5.0 percentage points), larger than our Poisson DID estimates. Panel B shows that the NTEIP initiatives reduce the likelihood of daytime accidents by 6.7% (3.1 out of 46.4 percentage points) and injuries by 9.3% (3.3 out of 35.3 percentage points), accounting for approximately 30% and 50% of the reductions in traffic accidents and injuries, respectively. Overall, our findings are robust to the use of the CSDID.

# 6 Conclusion

For this study, we used administrative data on traffic accidents and a staggered DID design to estimate how the NTEIP measures implemented on a large scale in Taipei's lanes and alleys since 2015 have impacted traffic safety. We focused on the improvement measures related to traffic accidents within the plan: the implementation of green-painted pedestrian walkways, adjustments to red/yellow no-parking lines, and the installation of speed limit and stop/slow signs.

Our estimates suggest that these NTEIP measures led to a 5% decrease in traffic accidents and an 8% reduction in daytime injuries. On average, there were 1.101 traffic accidents and 0.873 people injured per year in each of the 3,414 treated alleys. Therefore, the NTEIP improvements were estimated to reduce the number of traffic injuries by about 188 accidents ($1.101 \times 3,414 \times 0.05 = 188$) and 238 people ($0.873 \times 3,414 \times 0.08 = 238$) per year. Interestingly, the percentage reduction in injuries attributed to the program was higher than that of accidents. This suggests that the program measures have reduced not only the likelihood of traffic accidents but also their severity.

While the NTEIP has effectively reduced daytime traffic accidents and injuries, its impact on nighttime incidents is not as pronounced. A plausible explanation for this is the visibility factor: during daylight hours, with ample illumination, road markings or signs introduced by the program are more easily spotted by road users. This visibility serves as a reminder for them to be more cautious about road conditions. In contrast, at night, insufficient lighting may prevent drivers from clearly identifying these signs and markings, thus limiting their effectiveness. Therefore, we recommend that any road markings or signs be adequately illuminated at night to insure their clear visibility. This would help alert and guide nighttime road users and thus improve road safety.

Finally, our research indicates that cost-effective solutions like painting pedestrian walkways green can significantly enhance traffic safety, even in densely populated areas with high levels of motorcycle traffic such as Taipei. Cities facing traffic challenges may benefit from adopting similar policies to improve neighborhood traffic safety.

Figure 1: Implementation of the NTEIP

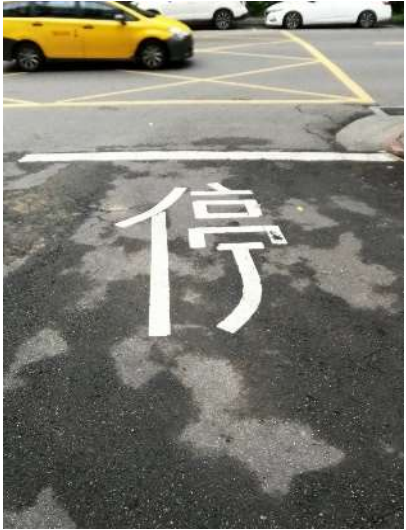

(a) Stop sign – Lane 92, Sec. 2, Xiuming Rd.

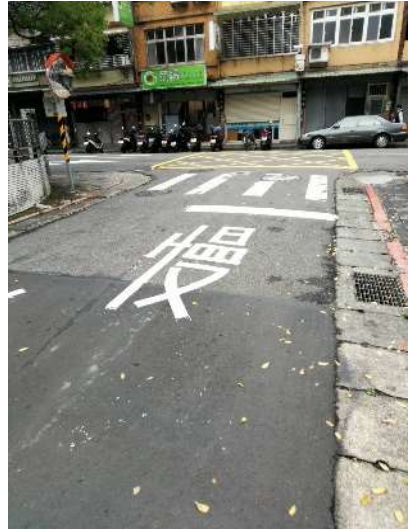

(b) Slow sign – Lane 66, Sec. 1, Xinguan Rd.

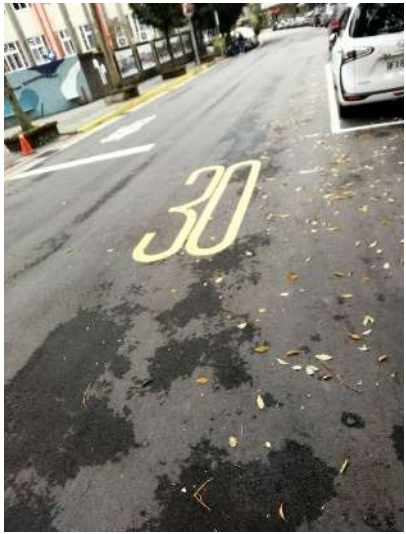

(c) Speed limit 30 sign – Lane 112, Sec. 2, Xiuming Rd

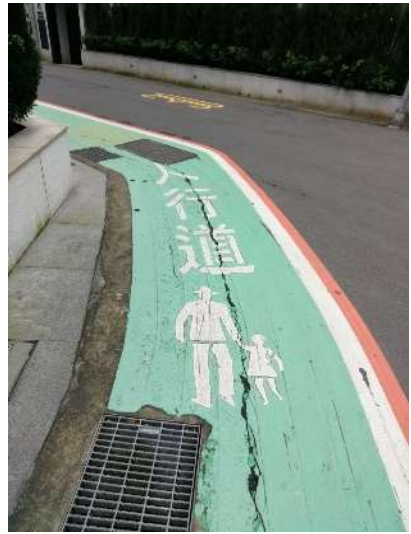

(d) Green-painted walkway – Lane 45, Sec. 2, Zhinan Rd.

Table 1: NTEIP – Road Classification

| Status Prior to NTEIP | Planned Improvements | Road Completeness | Classification |
|---|---|---|---|
| Implementation completed | No | Complete | Always treated |
| Partially or not implemented | After planning, implement measures other than green-painted walkways, red/yellow no-parking lines, and/or speed limit and stop/slow signs | Complete | Always treated |
| Partially or not implemented | After planning, implement green-painted walkways, red/yellow no-parking lines, and/or speed limit and stop/slow signs | Complete | Treatment group |
| Partially or not implemented | No | Incomplete | Never treated |

Note: The roads planned under the NTEIP are divided into three categories: the always-treated group, the treatment group and the never-treated group, according to the road implementation status at the time of program planning, the improvement content of the plan, and the final road completeness.

Table 2: Road Naming Principles in Taipei

| Road Width | Road Length | Name |
|---|---|---|
| 15 meters or more (25 meters or more) | Roads longer than 300 meters will be segmented | Road (Avenue) |
| More than 8 meters but less than 15 meters | | Street |
| The above conditions are not met | | Lane |
| Small passage in the lane | | Alley |

Note: According to the naming rules of the "Taipei City Road Naming and House Numbering Autonomy Ordinance," roads are named differently based on their width and length.

Table 3: Summary Statistics

| Road Type | Number of Road | Time | RTA | Injured |
|---|---|---|---|---|
| Always-treated group | 1,891 | All-Mean | 1.993 | 1.510 |
| | | | (3.441) | (2.877) |
| | | Day-Mean | 1.384 | 1.043 |
| | | | (2.405) | (2.065) |
| | | Night-Mean | 0.609 | 0.467 |
| | | | (1.351) | (1.207) |
| Treated group | 3,414 | All-Mean | 1.557 | 1.226 |
| | | | (2.593) | (2.418) |
| | | Day-Mean | 1.105 | 0.875 |
| | | | (1.894) | (1.807) |
| | | Night-Mean | 0.452 | 0.351 |
| | | | (0.996) | (0.978) |
| Never-treated group | 1,551 | All-Mean | 0.645 | 0.492 |
| | | | (1.463) | (1.364) |
| | | Day-Mean | 0.466 | 0.354 |
| | | | (1.121) | (1.049) |
| | | Night-Mean | 0.179 | 0.138 |
| | | | (0.554) | (0.566) |

Note: This table compiles data from a total of 6,856 lanes and alleys under the NTEIP from 2013 to 2020. It displays the mean and standard deviation for the number of traffic accidents and injuries, both overall and separately for day and night.

Table 4: Main Results

|  | (1) | (2) | (4) | (5) |
|---|---|---|---|---|
| VARIABLES | RTA | RTA | Injured | Injured |
| D | -0.0351 | -0.0339 | -0.0655** | -0.0698** |
|  | (0.0236) | (0.0236) | (0.0321) | (0.0322) |
| Pop_density |  | 0.00109 |  | -0.145 |
|  |  | (0.438) |  | (0.597) |
| Veh_density |  | -0.848** |  | -0.801* |
|  |  | (0.366) |  | (0.476) |
| Education |  | 0.0401*** |  | -0.0307 |
|  |  | (0.0148) |  | (0.0228) |
| Observations | 24,832 | 24,832 | 21,968 | 21,968 |
| Number of roads | 3,104 | 3,104 | 2,746 | 2,746 |
| Road FE | YES | YES | YES | YES |
| Year FE | YES | YES | YES | YES |

Note: This table estimates the effects of NTEIP on the number of traffic accidents and injuries. All models were estimated using Poisson regression, and the coefficients reported are the percentage changes. For example, a coefficient of 0.1 indicates a 10% increase in accidents or injuries. Data are from 3,414 treated roads from 2012 to 2019. Standard errors are clustered by roads. ***, ** and * represent significance levels of 1%, 5% and 10%, respectively.

Table 5: Number of Daytime Traffic Accidents and Injuries

|  | (1) Day RTA | (2) Day RTA | (3) Day Injured | (4) Day Injured |
|---|---|---|---|---|
| VARIABLES |  |  |  |  |
| D | -0.0490* | -0.0476* | -0.0815** | -0.0890** |
|  | (0.0259) | (0.0258) | (0.0352) | (0.0352) |
| Pop_density |  | -0.0558 |  | -0.957 |
|  |  | (0.490) |  | (0.652) |
| Veh_density |  | -0.439 |  | -0.0755 |
|  |  | (0.404) |  | (0.543) |
| Education |  | 0.0404** |  | -0.0297 |
|  |  | (0.0173) |  | (0.0259) |
| Observations | 23,976 | 23,976 | 20,768 | 20,768 |
| Number of Roads | 2,997 | 2,997 | 2,596 | 2,596 |
| Road FE | YES | YES | YES | YES |
| Year FE | YES | YES | YES | YES |

Note: This table estimates the effects of the NTEIP on the number of traffic accidents and injuries occurring in daytime. All models were estimated using Poisson regression, and the coefficients reported are the percentage changes. For example, a coefficient of 0.1 indicates a 10% increase in accidents or injuries. Data are from 3,414 treated roads from 2012 to 2019. Standard errors are clustered by roads. ***, ** and * represent significance levels of 1%, 5% and 10%, respectively.

Table 6: Number of Nighttime Traffic Accidents and Injuries

| VARIABLES | (1) Night RTA | (2) Night RTA | (3) Night Injured | (4) Night Injured |
|---|---|---|---|---|
| D | 0.00656 | 0.00699 | -0.0151 | -0.0115 |
|  | (0.0420) | (0.0421) | (0.0593) | (0.0593) |
| Pop_density |  | 0.0599 |  | 1.854* |
|  |  | (0.703) |  | (1.045) |
| Veh_density |  | -1.816*** |  | -2.611*** |
|  |  | (0.595) |  | (0.844) |
| Education |  | 0.0421 |  | -0.0327 |
|  |  | (0.0269) |  | (0.0435) |
| Observations | 20,088 | 20,088 | 15,480 | 15,480 |
| Number of Roads | 2,511 | 2,511 | 1,935 | 1,935 |
| Road FE | YES | YES | YES | YES |
| Year FE | YES | YES | YES | YES |

Note: This table estimates the effects of the NTEIP on the number of traffic accidents and injuries occurring in nighttime. All models were estimated using Poisson regression, and the coefficients reported are the percentage changes. For example, a coefficient of 0.1 indicates a 10% increase in accidents or injuries. Data are from 3,414 treated roads from 2012 to 2019. Standard errors are clustered by roads. ***, ** and * represent significance levels of 1%, 5% and 10%, respectively.

Table 7: Dynamic Effects of NTEIP

| VARIABLES | (1) RTA | (2) Injured | (3) Day RTA | (4) Day Injured | (5) Night RTA | (6) Night Injured |
|---|---|---|---|---|---|---|
| $D_{-3}$ | 0.0127 | 0.0208 | 0.0395 | 0.0741 | -0.0476 | -0.0965 |
|  | (0.0398) | (0.0540) | (0.0448) | (0.0629) | (0.0718) | (0.0979) |
| $D_{-2}$ | 0.0184 | 0.0288 | 0.0281 | 0.0334 | -0.00274 | 0.0254 |
|  | (0.0260) | (0.0358) | (0.0302) | (0.0415) | (0.0468) | (0.0671) |
| $D_0$ | -0.0102 | -0.0174 | -0.0208 | -0.0405 | 0.0179 | 0.0502 |
|  | (0.0275) | (0.0384) | (0.0317) | (0.0439) | (0.0473) | (0.0665) |
| $D_1$ | -0.0486 | -0.0943* | -0.0852* | -0.156*** | 0.0511 | 0.0782 |
|  | (0.0384) | (0.0529) | (0.0436) | (0.0603) | (0.0635) | (0.0868) |
| $D_2$ | -0.0566 | -0.0951 | -0.0716 | -0.116 | -0.0154 | -0.0308 |
|  | (0.0544) | (0.0738) | (0.0611) | (0.0844) | (0.0865) | (0.119) |
| $D_3$ | -0.0867 | -0.133 | -0.127 | -0.178 | 0.0261 | 0.00403 |
|  | (0.0719) | (0.0982) | (0.0814) | (0.114) | (0.111) | (0.152) |
| Pop_density | -0.136 | -0.267 | -0.147 | -0.872 | -0.173 | 1.222 |
|  | (0.451) | (0.627) | (0.512) | (0.692) | (0.714) | (1.082) |
| Veh_density | -0.776** | -0.746 | -0.404 | -0.154 | -1.661*** | -2.222*** |
|  | (0.355) | (0.466) | (0.399) | (0.545) | (0.579) | (0.820) |
| Education | 0.0404*** | -0.0307 | 0.0410** | -0.0299 | 0.0413 | -0.0333 |
|  | (0.0149) | (0.0228) | (0.0173) | (0.0259) | (0.0267) | (0.0431) |
| Observations | 24,832 | 21,968 | 23,976 | 20,768 | 20,088 | 15,480 |
| Number of Roads | 3,104 | 2,746 | 2,997 | 2,596 | 2,511 | 1,935 |
| Road FE | YES | YES | YES | YES | YES | YES |
| Year FE | YES | YES | YES | YES | YES | YES |

Note: This table estimates the dynamic effects of the NTEIP on the number of traffic accidents and injuries, both overall and separately for day and night. All models were estimated using Poisson regression, and the coefficients reported are percentage changes. For example, a coefficient of 0.1 indicates a 10% increase in accidents or injuries. Data are from 3,414 treated roads from 2012 to 2019. Standard errors are clustered by roads. ***, ** and * represent significance levels of 1%, 5% and 10%, respectively

Table 8: Effects of Specific Improving Content

| VARIABLES | (1) RTA | (2) Injured | (3) Day RTA | (4) Day Injured | (5) Night RTA | (6) Night Injured |
|---|---|---|---|---|---|---|
| D_Sidewalk | -0.0283 | -0.0817 | -0.0633 | -0.111* | 0.0609 | 0.000510 |
|  | (0.0412) | (0.0540) | (0.0416) | (0.0604) | (0.0787) | (0.0988) |
| D_RedYellow | -0.0109 | -0.0214 | -0.00290 | -0.0261 | -0.0261 | -0.00286 |
|  | (0.0230) | (0.0309) | (0.0247) | (0.0332) | (0.0414) | (0.0586) |
| D_Speed | 0.0207 | -0.00702 | -0.00610 | -0.0453 | 0.0881* | 0.0904 |
|  | (0.0264) | (0.0385) | (0.0300) | (0.0425) | (0.0468) | (0.0709) |
| Pop_density | -0.0775 | -0.167 | -0.0315 | -0.877 | -0.237 | 1.586 |
|  | (0.445) | (0.604) | (0.499) | (0.661) | (0.712) | (1.059) |
| Veh_density | -0.810** | -0.766 | -0.426 | -0.0704 | -1.731*** | -2.511*** |
|  | (0.362) | (0.471) | (0.401) | (0.540) | (0.584) | (0.831) |
| Education | 0.0407*** | -0.0294 | 0.0409** | -0.0283 | 0.0439 | -0.0310 |
|  | (0.0149) | (0.0228) | (0.0173) | (0.0259) | (0.0267) | (0.0432) |
|  |  |  |  |  |  |  |
| Observations | 24,832 | 21,968 | 23,976 | 20,768 | 20,088 | 15,480 |
| Number of Roads | 3,104 | 2,746 | 2,997 | 2,596 | 2,511 | 1,935 |
| Road FE | YES | YES | YES | YES | YES | YES |
| Year FE | YES | YES | YES | YES | YES | YES |

Note: This table estimates the effects of each NTEIP enhancement measure (green-painted walkways, no-parking red/yellow lines, and speed limit, stop, and/or slow signs) on the number of traffic accidents and injuries, both overall and separately for day and night. All models were estimated using Poisson regression, and the coefficients reported represent the rate of change. For example, a coefficient of 0.1 indicates a 10% increase in accidents or injuries. Data are from the 3,414 treated samples from 2012 to 2019. Standard errors are clustered by roads. ***, **, and * represent significance levels of 1%, 5%, and 10%, respectively.

Table 9: Robustness Check - A1/A2 and A3 Number of Accidents

|  | (1) | (2) | (3) | (4) | (5) | (6) |
|---|---|---|---|---|---|---|
|  |  | Day | Night |  | Day | Night |
| VARIABLES | A1/A2 | A1/A2 | A1/A2 | A3 | A3 | A3 |
| D | -0.0646** | -0.0702** | -0.0415 | 0.0188 | -0.00665 | 0.0791 |
|  | (0.0296) | (0.0324) | (0.0565) | (0.0353) | (0.0416) | (0.0643) |
| Pop_density | -0.415 | -0.981 | 1.014 | 0.969 | 1.717** | -0.947 |
|  | (0.554) | (0.606) | (0.931) | (0.619) | (0.716) | (1.034) |
| Veh_density | -0.712 | -0.102 | -2.286*** | -1.214** | -1.132* | -1.339 |
|  | (0.462) | (0.509) | (0.789) | (0.507) | (0.606) | (0.818) |
| Education | -0.00720 | -0.00359 | -0.0153 | 0.0944*** | 0.0951*** | 0.0974*** |
|  | (0.0213) | (0.0241) | (0.0397) | (0.0200) | (0.0231) | (0.0365) |
|  |  |  |  |  |  |  |
| Observations | 21,968 | 20,768 | 15,480 | 22,576 | 20,688 | 15,568 |
| Number of Roads | 2,746 | 2,596 | 1,935 | 2,822 | 2,586 | 1,946 |
| Road FE | YES | YES | YES | YES | YES | YES |
| Year FE | YES | YES | YES | YES | YES | YES |

Note: This table estimates the effects of the NTEIP on the number of A1/A2 and A3 traffic accidents, both overall and separately for day and night. All models were estimated using Poisson regression, and the coefficients reported represent percentage changes. For example, a coefficient of 0.1 indicates a 10% increase in accidents or injuries. Data are from the 3,414 treated samples from 2012 to 2019. Standard errors are clustered by roads. ***, **, and * represent significance levels of 1%, 5%, and 10%, respectively.

Table 10: Robustness Check – Callaway and Sant'Anna's (2021) Estimator

|  | (1) RTA | (2) Injured | (3) Day RTA | (4) Day Injured | (5) Night RTA | (6) Night Injured |
|---|---|---|---|---|---|---|
| Panel A: *Normalized by Village Mean* | | | | | | |
| D | -0.010* | -0.011* | -0.011** | -0.009** | 0.001 | -0.003 |
|  | (0.006) | (0.006) | (0.005) | (0.005) | (0.003) | (0.004) |
| Mean in 2015 | 0.071 | 0.069 | 0.051 | 0.050 | 0.020 | 0.019 |
| | | | | | | |
| Panel B: *Extensive Margin* | | | | | | |
| D | -0.056*** | -0.049*** | -0.031* | -0.031* | -0.015 | -0.013 |
|  | (0.018) | (0.018) | (0.019) | (0.018) | (0.018) | (0.015) |
| Mean in 2015 | 0.544 | 0.396 | 0.463 | 0.333 | 0.264 | 0.172 |
| | | | | | | |
| Road FE | YES | YES | YES | YES | YES | YES |
| Year FE | YES | YES | YES | YES | YES | YES |
| Controls | YES | YES | YES | YES | YES | YES |

Note: All models were estimated using the estimator developed by Callaway and Sant'Anna (2021). Control variables are population density, vehicle density, and education level at the district level. All models are estimated from the treated sample from 2012 to 2019. ***, ** and * represent significance levels of 1%, 5% and 10%, respectively.

# Appendix

## Description of Road Classification

This paper classified the roads that were planned to receive improvements under the NTEIP into an always-treated group, a treatment group, and a never-treated group; this classification also affected the setting of program dummy variables in the model. To illustrate how each road is classified, Table A1 presents some data from Chongqing Village in Datong District as an example, and adds the results of the road classifications used in this article in the last column.

Road 1 is classified as part of the treatment group; Road 2, having had improvements already implemented at the time of planning, does not require additional improvements under the NTEIP, and is thus classified as part of the always-treated group. The improvement measures for Road 3 cannot be implemented, and as the road is incomplete under the NTEIP, it is classified as part of the never-treated group. Road 4, although planned for added motorcycle parking, does not need any of the main improvement measures analyzed in this paper to be installed (i.e., green-painted pedestrian walkways, red/yellow lines no-parking, speed limit and stop/slow signs), and is therefore classified as part of the always-treated group. Road 5, with its planned installation of green-painted pedestrian walkways, is categorized as part of the treatment group.

Table A1: Examples of road classification in Chongqing Village, Datong District

| Road | Improving Content And Location | Car Parking Space | Motorcycle Parking Space Vertical | Parallel | Marked Walkway | Length | No-Parking Red Line | Length | No-Parking Yellow Line | Length | Motorcycle Exit for Arcades | Speed Limit 30 Signs | Stop & Slow-down Signs | Bicycle Parking Space | Completeness | Classification |
|---|---|---|---|---|---|---|---|---|---|---|---|---|---|---|---|---|
| 1 | 1. Set stop sign and stop line 2. There is no slow sign in the current location 3. There is a stop sign in the current location | | | | | | | | | | | | V stop | | V | Treatment group |
| 2 | Completed | | | | | | | | | | | | | | V | Always Treated |
| 3 | Cannot be implemented | | | | | | | | | | | | | | X | Never treated |
| 4 | On the west side, add 3 motorcycle parking spaces | | 3 | | | | | | | | | | | | V | Always Treated |
| 5 | 1. Add a green-painted walkway of about 35 meters on the west side 2. Add 17 motorcycle parking spaces on the east side | | 17 | | V | 35.0 | | | | | | | | | V | Treatment group |

Note: This information is from of the NTEIP section of the Taipei City Government's Department of Transportation, showing the planning status of some roads in Chongqing Village, Datong District and the improvements planned for each road to illustrate how the planned improvements are used to classify roads in this article (always-treated, treatment, and never-treated groups).